# Dependence of quantum correlations of twin beams on pump finesse of optical parametric oscillator

**Dong Wang, Yana Shang, Xiaojun Jia, Changde Xie[*], and Kunchi Peng**

**State key Laboratory of Quantum Optics and Quantum Optics Devices, Institute of Opto-Electronics, Shanxi University, Taiyuan 030006, China**

The dependence of quantum correlation of twin beams on the pump finesse of an optical parametric oscillator is studied with a semi-classical analysis. It is found that the phase-sum correlation of the output signal and idler beams from an optical parametric oscillator operating above threshold depends on the finesse of the pump field when the spurious pump phase noise generated inside the optical cavity and the excess noise of the input pump field are involved in the Langevin equations. The theoretical calculations can explain the previously experimental results, quantitatively.

## I INTRODUCTION

As an important device in nonlinear optics, quantum optics and quantum information, the optical parametric oscillator (OPO) has been extensively studied and applied since 1960s. Especially, it has become one of the most successful tools for the generation of entangled states of light in continuous variable (CV) quantum information systems [1]. Early, Reid and Drummond theoretically demonstrated that the Einstein-Podolsky-Rosen (EPR) entangled states can be generated from a nondegenerate OPO (NOPO) operating both above and below its threshold [2-5]. For the first time, CV EPR entanglement was experimentally realized by Ou et al. with a NOPO below threshold in 1992 [6]. In recent years, the optical CV entangled states with quantum correlations of amplitude and phase quadratures of light fields produced from OPOs or NOPOs below threshold have been used in quantum information systems to realize the unconditional quantum teleportation [7], quantum dense coding [8], quantum entanglement swapping [9], quantum key distribution [10, 11] and a variety of quantum communication networks [12-14]. Although the intensity difference quantum correlations of twin beams from NOPO above threshold were measured experimentally and were effectively applied by several groups since the first experiment achieved by Heidmann et al. in 1987 [15-20], the phase correlations of them were not been observed up to 2005 owing to technical difficulty in measuring the phase noise of twin beams with nondegenerate frequencies. In 2005, Laurat et al. forced the NOPO to oscillate in a strict frequency-degenerate situation by inserting a $\lambda/4$ plate inside the optical cavity with a finesse of $\approx 102$ for the pump laser, and observed a 3 dB phase-sum variance above the shot noise limit (SNL) [21]. Later, 0.8dB phase correlation below the SNL between twin beams with different frequency from a NOPO for a pump power of $\sim 4\%$ above threshold was measured by Villar et al. by scanning a pair of tunable ring analysis cavities [22]. In the experiment of Ref. [22], when the pump power was higher than 1.07 times of the threshold, the phase-sum noise of twin beams was lager than that of the SNL and thus the quantum correlation of the phase quadratures disappeared. Successively, our group detected the phase-sum correlation of the twin beams with two sets of unbalanced Match-Zehnder interferometers [23]. In this experiment the phase-sum correlation of 1.05dB lower the SNL was recorded at a pump power of 230mW which was almost two times of the threshold of 120 mW. In 2006, the phase-sum correlation of 1.35dB below the SNL between twin beams with a stable frequency-difference was obtained with a doubly resonant NOPO without the resonance of the pump field [24].

To explain why the experimentally measured phase-sum correlations of twin beams were always lower than that predicted by theory and why it easily disappeared in some experimental systems, the influence of the excess noise of the pump field was theoretically and experimentally studied recently [25-28]. Especially it was discovered by Villar et al. [28] that the spurious pump noise is generated inside the OPO cavity containing a nonlinear crystal, even for a shot-noise limited input pump beam and without parametric oscillation. They analyzed the physical origin of this phenomenon and assumed that the pump phase noise generated inside cavity due to the effect of the intensity-dependent index of refraction should be mainly responsible to the lower phase-sum correlation. Thus, they pointed out that the phase shifts accumulated inside the cavity with a lower finesse of pump laser should be smaller, hence the spurious noise generated should also be smaller, probably. Very recently, we experimentally investigated the influence of the excess pump noise on the entanglement of twin beams by adding different excess phase noise on the input pump laser outside the cavity [29]. In this experiment, the noise spectra of the intensity-difference and the phase-sum of twin beams were measured at three analysis frequencies of 2MHz, 5MHz and 10MHz under three different pump phase noises. The experimental results showed that the measured phase-sum correlations were still worse than that calculated with the theoretical formula in which the excess pump phase noise was involved. We considered that is because the possibly spurious phase-noise of the pump laser produced inside the NOPO was not counted in the formula.

It has been proved that in the calculations of the quantum correlations between the output signal and idler from NOPO，the standard full quantum theory almost leads to the same results with that deduced with the semiclassical methods [30-33]. For conveniently comparing with experiments, in this paper, we present a semiclassical analysis of quantum correlations for the intensity-difference and the phase-sum of twin beams. A set of semiclassical Langevin equations involving the excess pump phase noise and the spurious phase noise produced inside cavity is given. By solving the Langevin equations the analytic expressions for the intensity-difference and the phase-sum noise spectra of twin beams are obtained. The expressions are compatible with that in Ref. [28, 30] if the excess pump phase noise and the spurious phase noise inside cavity are not considered. All physical parameters in the expressions are experimentally measurable parameters, thus we can conveniently compare the theoretical calculations and the experimental results. The numerical calculations based on the expressions of the noise spectra show that the phase-sum noise spectrum of twin beams depends on the finesse of the pump laser. Our calculations proved quantitatively the physical analysis on this phenomenon in Ref. [28]. The published experimental results in Ref. [21-25] can be fit reasonably to the theoretical results if the appropriate parameters charactering the spurious phase noise and the excess noise of input pump field are taken.

## II LANGEVIN EQUATIONS INVOLVING EXCESS PUMP NOISE AND INTRACAVITY SPURIOUS PHASE NOISE

The semiclassical motion equations for the pump mode $\alpha_0$, signal mode $\alpha_1$ and idler mode $\alpha_2$ inside a triple resonant NOPO can are described by Eq. (1),

$$\tau \dot{\alpha}_1 = -(\gamma + \mu)\alpha_1 + 2\chi \alpha_0 \alpha_2^* + \sqrt{2\gamma}\alpha_1^{in} + \sqrt{2\mu}\beta_1^{in}$$
$$\tau \dot{\alpha}_2 = -(\gamma + \mu)\alpha_2 + 2\chi \alpha_0 \alpha_1^* + \sqrt{2\gamma}\alpha_2^{in} + \sqrt{2\mu}\beta_2^{in}$$
$$\tau \dot{\alpha}_0 = -(\gamma_0 + \mu_0)\alpha_0 - 2\chi \alpha_1 \alpha_2 + \sqrt{2\gamma_0}\alpha_0^{in} + \sqrt{2\mu_0}\beta_0^{in}$$

(1)

which can be obtained by adding Gaussian white noise to classical electrodynamics [31]. In Eq. (1), $\tau$ is the round-trip time, which is assumed to be the same for all three fields. $\chi$ is the nonlinear coupling parameter. $\gamma_i$ and $\mu_i$ ($i = 0,1,2$) are the one pass losses associated with the coupling mirror of the cavity and with all other losses, respectively. Without losing generality, we assume that the losses of the signal and idler modes are balanced, thus we have $\gamma = \gamma_1 = \gamma_2$ and $\mu = \mu_1 = \mu_2$. $\alpha_i^{in}$ and $\beta_i^{in}$ are the incoming fields, associated with the coupling mirror and with the intracavity loss mechanism, respectively.

Solving Eq. (1), the stationary state values are obtained:

$$\bar{\alpha}_1^2 = \bar{\alpha}_2^2 = \frac{\gamma_0' \gamma'}{4\chi^2}(\sigma - 1)$$
$$\bar{\alpha}_0^2 = \frac{\gamma'^2}{4\chi^2} \tag{2}$$

where the loss parameters $\gamma' = \gamma + \mu$ and $\gamma_0' = \gamma_0 + \mu_0$. In the case above threshold, the pump parameter $\sigma$ is larger than 1:

$$\sigma = 2\sqrt{\frac{2\chi^2 \gamma_0}{\gamma_0'^2 \gamma'^2}} \bar{\alpha}_0^{in} \tag{3}$$

where $\bar{\alpha}_0^{in}$ stands for the mean amplitude of the input field.

In order to get the noise dynamic equations, a semiclassical method is used. We define the fluctuation operators $\delta\alpha_i$ and $a_i = \bar{\alpha}_i + \delta\alpha_i$, $\bar{\alpha}_i$ is the mean value of $a_i$. Introducing the real and imaginary parts of the field, we get the noise operators of the amplitude and phase quadratures:

$$\begin{aligned} p_i &= \delta\alpha_i + \delta\alpha_i^* \\ q_i &= -i(\delta\alpha_i - \delta\alpha_i^*) \end{aligned} \quad (i = 0,1,2) \tag{4}$$

It has been well-known that the amplitude quadratures of the output twin beams are correlated and their phase quadratures are anticorrelated, respectively [30]. The amplitude-difference and the phase-sum noise operators of the twin beams are expressed by:

$$p = \frac{1}{\sqrt{2}}(p_1 - p_2)$$
$$q = \frac{1}{\sqrt{2}}(q_1 + q_2)$$
(5)

From Eq. (1) and using the input and output relation:

$$p^{out}(\omega) = \sqrt{2\gamma}\, p(\omega) - p^{in}(\omega) \tag{6}$$

we obtain the correlation spectrum $p^{out}(\omega)$ of the amplitude-difference:

$$p^{out}(\omega) = \frac{1}{2\gamma' + i\omega\tau}[\sqrt{2\gamma}\, p^{in}(\omega) + p'^{in}(\omega)] \tag{7}$$

where ω is the analysis frequency; $p^{in}(\omega)$ and $p'^{in}(\omega)$ are the vacuum noises associated with the cavity mirror and the intracavity loss respectively, both of which can be normalized to 1. We see that any parameter of pump mode is not involved in the right side of Eq. (7). That is to say, the amplitude-difference of the output twin beams doesn't depend on the pump intensity and the pump noise. The noise power spectrum of the amplitude-difference is given from Eq. (7):

$$S_p(\omega) = 1 - \frac{TT'}{T'^2 + \omega^2\tau^2} \tag{8}$$

where $T' = T + \delta$, $T = 2\gamma$ is the transmission coefficient of the output mirror and $\delta = 2\mu$ is the intracavity loss of twin beams in the NOPO. Eq. (8) is totally the same with the result deduced in Ref. [30] which has been extensively applied.

However, for the phase-sum we have to consider the influence of the pump noises since it can not be eliminated. It has been pointed out in Ref. [28] that the phase noise of the pump field in a NOPO with a nonlinear crystal will increase. Thus the crystal in an optical cavity can be regarded as a gain medium for the phase noise of the pump field [28]. We introduce a gain factor ε in the Langevin equation for the phase quadrature $q_0$ to characterize the effect of the spurious phase noise which is continuously gained in the crystal. Substituting Eqs. (4) and (5) into Eq. (1), we obtain the Langevin equations for the phase motion:

$$\tau \dot{q}_1 = -\gamma'(q_1+q_2)+\sqrt{\gamma_0'\gamma'(\sigma-1)}q_0+\sqrt{2\gamma}q_1^{in}+\sqrt{2\mu}q_{\beta 1}^{in}$$
$$\tau \dot{q}_2 = -\gamma'(q_2+q_1)+\sqrt{\gamma_0'\gamma'(\sigma-1)}q_0+\sqrt{2\gamma}q_2^{in}+\sqrt{2\mu}q_{\beta 2}^{in} \qquad (9)$$
$$\tau \dot{q}_0 = -\gamma_0'q_0+\varepsilon q_0-\sqrt{\gamma_0'\gamma'(\sigma-1)}(q_1+q_2)+\sqrt{2\gamma_0}q_0^{in}+\sqrt{2\mu_0}q_{\beta 0}^{in}$$

where $q_i^{in}$ and $q_{\beta i}^{in}$ $(i=0,1,2)$ are the phase quadratures of the incoming fields associated with the cavity mirror and the intracavity loss mechanism respectively, both of which can be normalized to 1. Solving these equations, we get:

$$q^{out} = \frac{\sqrt{2\gamma}\sqrt{2}\sqrt{\gamma_0'\gamma'(\sigma-1)}(\sqrt{2\gamma_0}q_0^{in}+\sqrt{2\mu_0}q_{\beta 0}^{in})+(i\omega\tau+\gamma_0'-\varepsilon)2\sqrt{\gamma\mu}q_\beta^{in}}{2\gamma'\gamma_0'\sigma-\omega^2\tau^2-2\gamma'\varepsilon+i\omega\tau(\gamma_0'+2\gamma'-\varepsilon)}$$
$$+\frac{\left[i\omega\tau(2\gamma-2\gamma'-\gamma_0'+\varepsilon)+2\gamma_0'\gamma-2\gamma\varepsilon-2\gamma'\gamma_0'\sigma+\omega^2\tau^2+2\gamma'\varepsilon\right]q^{in}}{2\gamma'\gamma_0'\sigma-\omega^2\tau^2-2\gamma'\varepsilon+i\omega\tau(\gamma_0'+2\gamma'-\varepsilon)} \qquad (10)$$

Assuming the excess noise of the input pump field at frequency $\omega$ is $E(\omega)$, i.e., $\left\langle\left|\delta q^{in}(\omega)\right|^2\right\rangle=1+E(\omega)$, the noise power spectrum formula of the phase-sum is obtained:

$$S_q(\omega) = 1-\frac{TT'(T_0'^2+4\omega^2\tau^2)+4T\varepsilon(2T'\varepsilon-T'T_0'\sigma-T\varepsilon-\delta\varepsilon)}{(T'T_0'\sigma-2\omega^2\tau^2-2T'\varepsilon)^2+\omega^2\tau^2(T_0'+2T'-2\varepsilon)^2}$$
$$+\frac{2TT'T_0'T_0(\sigma-1)}{(T'T_0'\sigma-2\omega^2\tau^2-2T'\varepsilon)^2+\omega^2\tau^2(T_0'+2T'-2\varepsilon)^2}E(\omega) \qquad (11)$$

where $T_0'=T_0+\delta_0$, $T_0=2\gamma_0$ is the transmission coefficient of the input mirror of the NOPO and $\delta_0=2\mu_0$ is the intracavity loss of the pump laser in the NOPO. If there is no the spurious noise inside the cavity $(\varepsilon=0)$, Eq. (11) goes to:

$$S_q(\omega) = 1 - \frac{TT'T_0'^2 + 4TT'\omega^2\tau^2}{(T'T_0'\sigma - 2\omega^2\tau^2)^2 + \omega^2\tau^2(T_0' + 2T')^2}$$

$$+ \frac{2TT'T_0'T_0(\sigma - 1)}{(T'T_0'\sigma - 2\omega^2\tau^2)^2 + \omega^2\tau^2(T_0' + 2T')^2} E(\omega) \quad (12)$$

If the cavity finesse of the pump field is much lower than that of signals ($T_0' \gg T'$), Eq. (12) can be simplified as:

$$S_q(\omega) = 1 - \frac{TT'}{T'^2\sigma^2 + \omega^2\tau^2} + \frac{2TT'(\sigma - 1)}{T'^2\sigma^2 + \omega^2\tau^2} E(\omega) \quad (13)$$

which is the same with that in Ref. [29] where the spurious pump phase noise was not considered. If the pump light is an ideal coherent laser without the excess noise, i.e., $E(\omega) = 0$, Eq. (13) can be further simplified as:

$$S_q(\omega) = 1 - \frac{TT'}{T'^2\sigma^2 + \omega^2\tau^2}$$

(14)

This equation is totally equivalent to the Eq. (25) in Ref. [30] which was deduced under the condition without the pump excess phase noise and the intracavity spurious pump phase noise. Thus the Eq. (11) is a general formula which is compatible with that obtained under the specific requirements.

### III NUMERICAL ANALYSIS ON PHASE-SUM CORRELATION OF TWIN BEAMS

In practically experimental system, the efficiency of the detector is always imperfect. Accounting for the detection efficiency of $\eta < 1,$ the noise power spectrum Eq. (11) of the phase-sum becomes:

$$S_q(\omega) = 1 - \eta \frac{TT'(T_0'^2 + 4\omega^2\tau^2) + 4T\varepsilon(2T'\varepsilon - T'T_0'\sigma^2 - T\varepsilon - \delta\varepsilon)}{(T'T_0'\sigma - 2\omega^2\tau^2 - 2T'\varepsilon)^2 + \omega^2\tau^2(T_0' + 2T' - 2\varepsilon)^2}$$

$$+ \frac{2TT'T_0'T_0(\sigma - 1)}{(T'T_0'\sigma - 2\omega^2\tau^2 - 2T'\varepsilon)^2 + \omega^2\tau^2(T_0' + 2T' - 2\varepsilon)^2} E \quad (15)$$

Fig. 1-4 show the dependence of the phase-sum correlation on the finesse of the pump field under different pump parameters $\sigma$ (Fig. 1), different intracavity noise $\varepsilon$ (Fig. 2) and different excess pump noise E with $\varepsilon \neq 0$ (Fig. 3) and $\varepsilon = 0$ (Fig. 4), respectively. In the four figures, other system parameters are the same, where $T = 5\%$, $\delta = \delta_0 = 0.5\%$, $\eta = 90\%$ and $\omega\tau = 0.025$. From Fig. 1, we can see that for a given $\varepsilon$ ($\varepsilon = 0.02$), the phase-sum noise increases along with the the increase of the pump power even in the case without the excess pump noise ($E = 0$). For higher pump power ($\sigma = 1.3$), the quantum correlation of the phase-sum disappears, i.e., the phase-sum noises are larger than the normalized SNL, for those NOPOs with the finesse in a certain range (from finesse $F = 68$ to $F = 134$ in Fig. 1). The results can be used to explain the experimental phenomena in Ref. [22], in which a critical pump parameter for the phase-sum correlation was measured (see Fig. 6 for detail). Fig. 2 shows that the phase-sum noise increases when $\varepsilon$ increases ($E = 0$). For a given NOPO, the phase-sum correlation can not be observed if the intracavity spurious phase noise is too high ($\varepsilon = 0.04$ for example). In general NOPOs, the excess pump noise and the spurious noise exist simultaneously and both influence the phase-sum correlation of twin beams. Fig. 3 shows the dependences of the phase-sum noises on $\varepsilon$ and $E$. It is pointed out that both the excess noise from the input pump field and the spurious noise produced inside the cavity decrease the quantum correlation of the phase-sum between the output signal and idler modes. From Fig. 4, we can see that the influence of the excess pump phase noise on the phase-sum noise of twin beams monotonously degrades as the pump finesse increases if the intracavity spurious pump phase noise is not considered ($\varepsilon = 0$). The physical reason of the effect is that in NOPOs with low pump finesses, the transmission of the input mirror for the pump field is quite high, so the incoming phase noise together with the pump field is also larger if $E \neq 0$. Due to that the phase-sum noise depends on a variety of physical parameters of both pump field and subharmonic fields [see Eq. (15)], the dependence of the phase sum correlation on the finesse of the pump field is not identical for different NOPO. The function curves of the phase-sum noise versus the pump finesse will change if other cavity parameters are changed. Generally, there is a maximum on the function curves if $\varepsilon \neq 0$. At first the phase-sum noise increases when the pump finesse increases from zero due to the effect of the intracavity spurious noise. However, the intracavity intensity of the pump field is also raised when the pump finesse increases under a given pump power. Thus, the effective nonlinear conversion efficiency in the NOPO will be enhanced, which must result in the increase of the quantum correlation between signal and idler modes. When the positive effect increasing the intracavity intensity of the pump field is superior to the negative effect gaining the spurious noise, the phase-sum noise will start to decrease if the pump finesse continuously increases.

For larger $\varepsilon$ and $E$, the phase-sum noise significantly depends on the pump finesse (see curve i of Fig. 2, curves i and ii of Fig. 3). When $\varepsilon$ is smaller the function curves of the phase-sum noise versus the pump finesse are flatter (see curves ii and iii of Fig. 2, curves iii and iv of Fig. 3). Comparing curves i, ii and iii, iv in Fig. 3, it is obvious that the influence of the intracavity spurious noise ($\varepsilon$) to the dependence of the phase-sum noise on the pump finesse is stronger than the influence of the excess phase noise of the input pump field ($E$).

## IV COMPAREISION OF THEORETICAL CALCULATIONS AND PREVIOUS EXPERIMENTS

After considering the influence of $E$ and $\varepsilon$, the theoretical calculations based on the real system parameters can match with the experimental results if appropriate values of $\varepsilon$ and $E$ are selected. In Ref. [21], the phase-sum correlation was not observed. We estimate $E$ was higher in their system probably. If taking $\varepsilon = 0.015$ and $E = 5$, the function curve of the phase-sum noise versus the pump finesse according to the experimental parameters of Ref. [21] is shown in Fig. 5. The star symbol denotes the experimental result (also for Fig. 6, Fig. 7 and Fig. 8), where the pump finesse is about 102 and the normalized phase-sum noise is about 2.1 corresponding to 3.2 dB above the SNL. The function curves for the experimental system of Ref. [28] are drawn in Fig. 6. Since they experimentally proved that the excess noise of the input pump field can be neglected at the analysis frequency of $f = \omega/(2\pi) = 27 MHz$ [28], we take $E=0$ and $\varepsilon=0.06$. The curves i and ii of Fig. 6 correspond to $\sigma=1.28$ and $\sigma=1.06$ respectively according to their experimental measurements. Under the low pump power ($\sigma=1.06$), the phase-sum correlation always exists (All phase-sum noises are smaller than that of the SNL). But under the higher pump power ($\sigma=1.28$), the phase-sum noises are larger than that of the SNL in the range of the pump finesses from 17 to 76. The theoretical curves are perfectly matched with the experimental results. The star on curve ii corresponds to the critical pump power for the phase-sum correlation in their experimental system. In Ref. [24], the pump field did not resonate, so we can consider the pump finesse was very low (close to 1). Fig. 7 is drawn with the parameters of the system of Ref. [24] where $\varepsilon=0.06$ and $E=5.6$ are taken for matching the experimentally measured phase-sum noise of 0.69 corresponding to 1.6 dB below the SNL. For our experimental system of Ref. [29] with the low finesse of $\approx 12$, if taking $\varepsilon=0.005$ and $E=0.06$, the measured phase-sum noise of 0.75 ( 1.25 dB below the SNL) will perfectly match with the theoretical curve (see Fig. 8).

Although the values of $\varepsilon$ and $E$ in Fig. 5-8 are not experimentally measured, these estimated values are reasonable. At least these calculations tell us that the previous experimental results on twin-beam generations from NOPOs above threshold achieved different groups can be explained by means of a semiclassical theory if the intracavity spurious phase noise and the excess phase noise of the input pump field are involved in the Langevin equations.

## V CONCLUSION

By solving the semiclassical Langevin equations involving the intracavity spurious pump phase noise and the excess noise of the input pump field, we obtained the expressions of the intensity-difference and the phase-sum noise spectra between the output signal and idler modes from a NOPO above threshold. The phase-sum quantum correlation of twin beams not only depends on the cavity parameters of the subharmonic field, but also depends on the finesse and the noises of the pump field. Especially, the phase-sum noise significantly increases when the spurious pump noise produced inside the cavity with a nonlinear crystal is higher. The dependence of the phase-sum correlation of twin beans on the system parameters of NOPO is more complex. Our calculations provide useful reference for the design of NOPO serving as a source of optical entanglement states. The expressions of the noise spectra presented in this paper are compatible with that obtained previously under the condition without considering the pump noises if taking $\varepsilon = 0$ and $E = 0$. Using the extended expressions, the previously experimental results can be reasonably explained if appropriate parameters characterizing pump noises are applied.

The NOPO above threshold is a helpful device to produce bright tunable entanglement optical beams which could be used to transfer quantum information from one frequency to another and to implement the quantum memory. The entanglement of twin beams with directly detectable intensity can be measured with a pair of analysis cavities [22], or unbalance M-Z interferometers [23, 29] without the need of a local oscillator, thus it might be conveniently applied to realize the quantum key distribution protocols based on entanglement states of light [11, 34, 35]. Clearing the excess pump phase noise, minimizing the intracavity spurious phase noise of the pump field and selecting appropriate parameters of optical cavity are the key factors for obtaining twin beams with higher phase-sum correlation.

## ACKNOWLEGEMENT

This research was supported by the PCSIRT (Grant No. IRT0516), Natural Science Foundation of China (Grants No. 60608012, 60608012, 60736040 and 10674088). *Correspondence should be addressed to Changde Xie: changde@sxu.edu.cn.

Figure captions:

FIG. 1 Phase-sum noise vs pump finesse with different pump parameters $\sigma = 1.3$ (i), $\sigma = 1.2$ (ii) and $\sigma = 1.1$ (iii); $E = 0$, $\varepsilon = 0.02$; other parameters: $T = 5\%$, $\delta = \delta_0 = 0.5\%$, $\eta = 90\%$ and $\omega\tau = 0.025$.

FIG. 2 Phase-sum noise vs pump finesse with different spurious phase noise, $\varepsilon = 0.04$ (i), $\varepsilon = 0.02$ (ii) and $\varepsilon = 0.01$ (iii); $E = 0, \sigma = 1.1$; other parameters are the same with that in FIG. 1.

FIG. 3 Phase-sum noise vs pump finesse with different $\varepsilon$ and $E$, $\varepsilon = 0.03$ and $E = 4$ (i), $\varepsilon = 0.03$ and $E = 2$ (ii), $\varepsilon = 0.01$ and $E = 4$ (iii), $\varepsilon = 0.01$ and $E = 2$ (iv); $\sigma = 1.1$; other parameters are the same with that in FIG. 1.

FIG. 4 Phase-sum noise vs pump finesse with different excess pump phase noise, $E = 8$ (i), $E = 6$ (ii), $E = 4$ (iii) and $E = 2$ (iv); $\varepsilon = 0, \sigma = 1.1$; other parameters are the same with that in FIG. 1.

FIG. 5 Phase-sum noise vs pump finesse for matching the experimental value of Ref. [21]. ( $T = 5\%$, $\delta = \delta_0 = 1\%$, $\omega\tau = 0.0099$, $f = \omega/(2\pi) = 5MHz$, $\eta = 90\%$, $\sigma = 1.1$, $\varepsilon = 0.15$, $E = 5$ )

FIG. 6 Phase-sum noise vs pump finesse for matching the experimental value of Ref. [28]. ( $T = 4\%$, $\delta = \delta_0 = 1\%$, $\omega\tau = 0.0512$, $f = \omega/(2\pi) = 27MHz$, $\eta = 80\%$,

$\varepsilon = 0.06$, $E = 0$ )

FIG. 7 Phase-sum noise vs pump finesse for matching the experimental value of Ref. [24]. ( $T = 1.8\%$, $\delta = \delta_0 = 1\%$, $\omega\tau = 0.0064$, $f = \omega/(2\pi) = 1.7MHz$, $\eta = 85\%$, $\varepsilon = 0.06$, $E = 5.6$ )

FIG. 8 Phase-sum noise vs pump finesse for matching the experimental value of Ref. [28]. ( $T = 3.2\%$, $\delta = \delta_0 = 1\%$, $\omega\tau = 0.0124$, $f = \omega/(2\pi) = 5MHz$, $\eta = 55\%$, $\varepsilon = 0.005$, $E = 0.06$ )

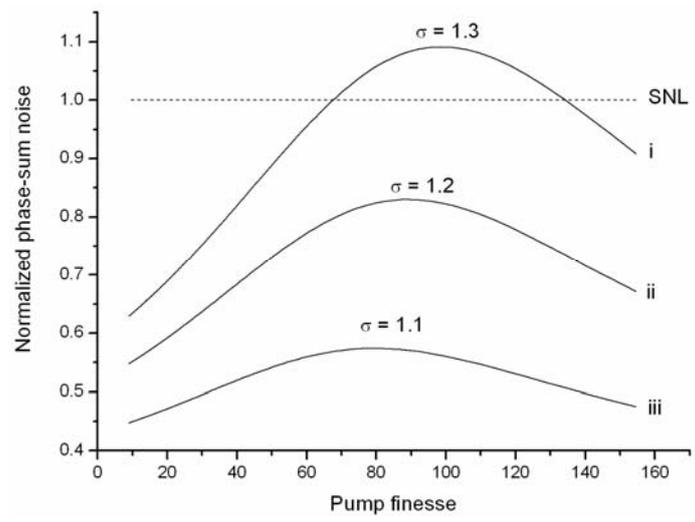

Figure 1

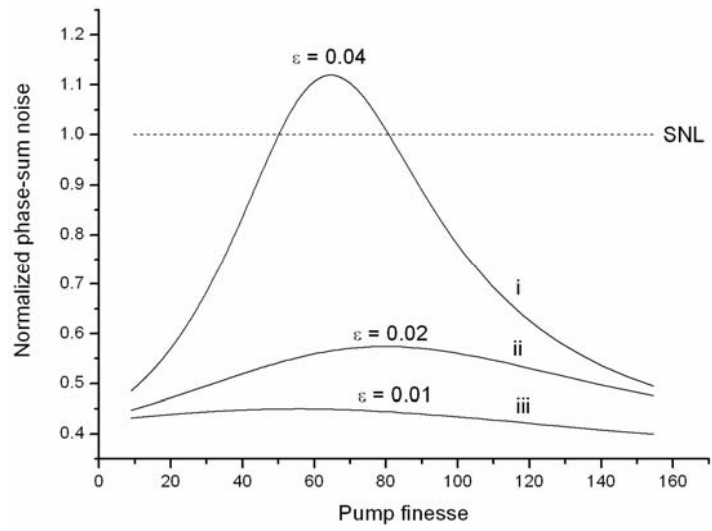

Figure 2

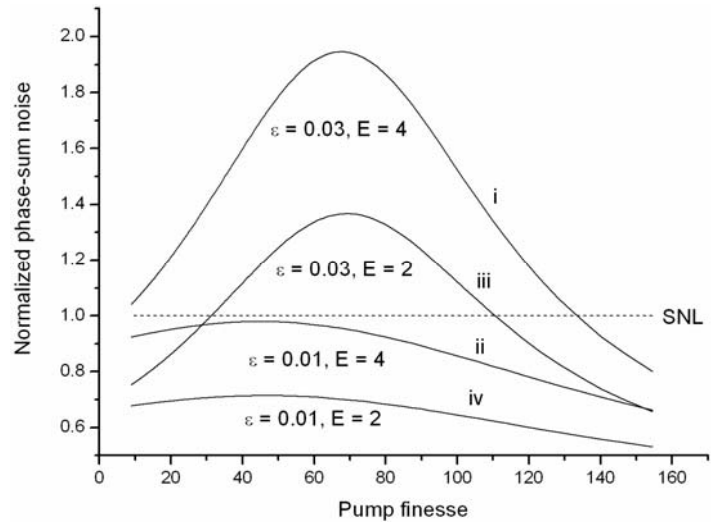

Figure 3

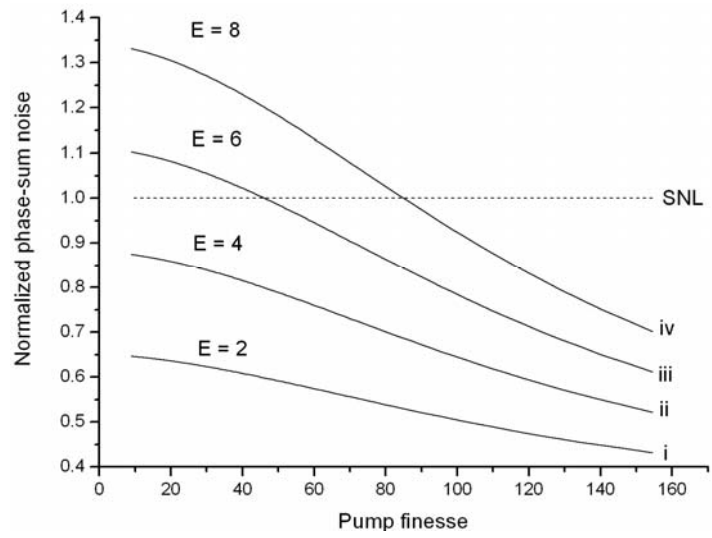

Figure 4

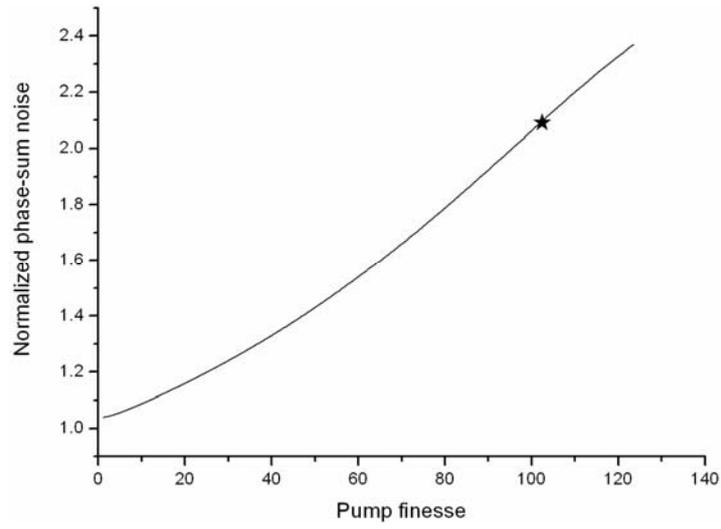

Figure 5

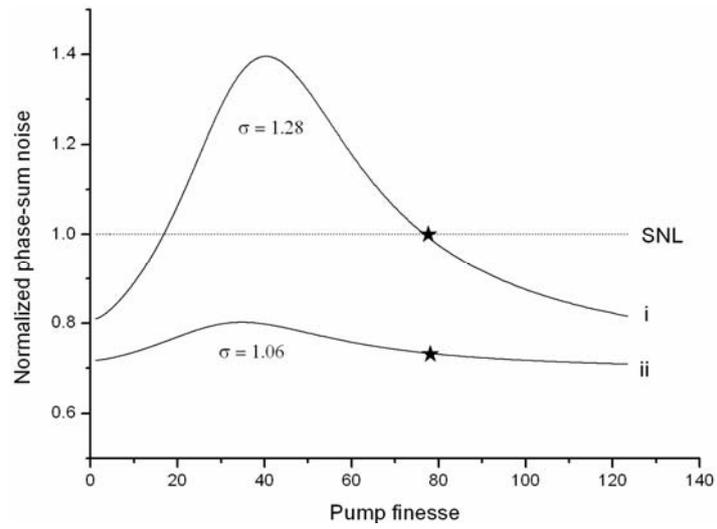

Figure 6

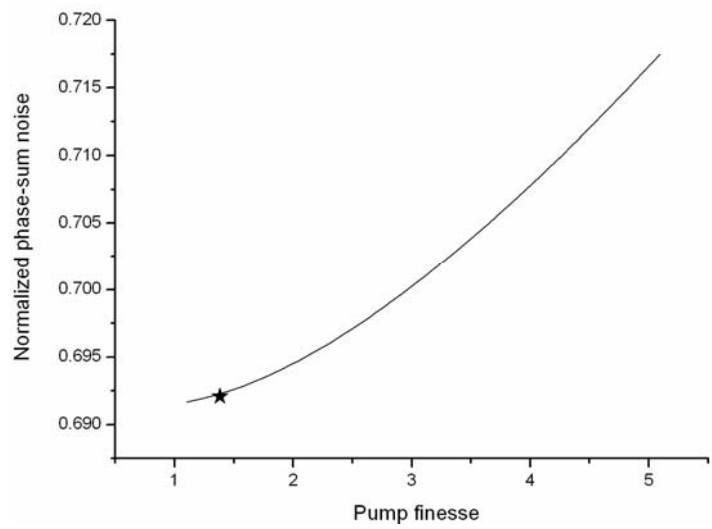

Figure 7

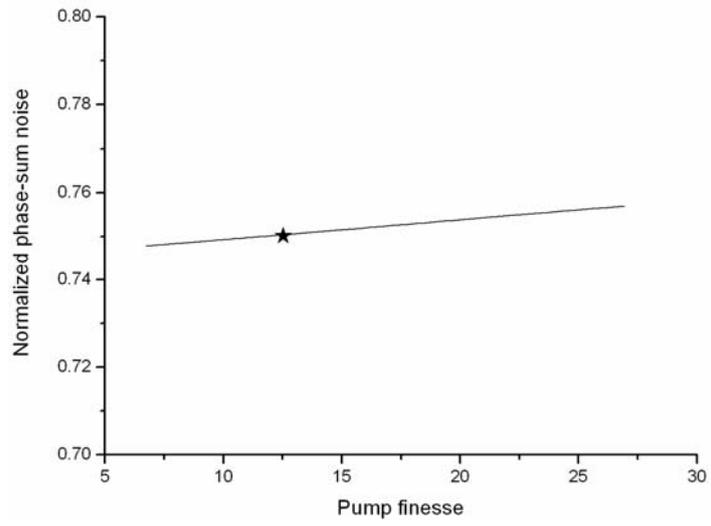

Figure 8